\documentclass[aps,prl,prabib,twocolumn,showpacs,superscriptaddress]{revtex4}
\usepackage{color}
\usepackage{graphicx}
\usepackage{amsmath}
\usepackage{amssymb}
\usepackage{amsfonts}
\usepackage{bm}

\begin{document}

\title{Seeing zeros of random polynomials:
quantized vortices in the ideal Bose gas}


\author{Yvan Castin}
\affiliation{Laboratoire Kastler Brossel, Ecole normale
sup\'erieure, 24 rue Lhomond, 75005 Paris, France}

\author{Zoran Hadzibabic}
\affiliation{Laboratoire Kastler Brossel, Ecole normale
sup\'erieure, 24 rue Lhomond, 75005 Paris, France}

\author{Sabine Stock}
\affiliation{Laboratoire Kastler Brossel, Ecole normale
sup\'erieure, 24 rue Lhomond, 75005 Paris, France}

\author{Jean Dalibard}
\affiliation{Laboratoire Kastler Brossel, Ecole normale
sup\'erieure, 24 rue Lhomond, 75005 Paris, France}

\author{Sandro Stringari}
\affiliation{Laboratoire Kastler Brossel, Ecole normale
sup\'erieure, 24 rue Lhomond, 75005 Paris, France}
 \affiliation{Coll\`ege de France, 11 place Marcelin Berthelot,
75005 Paris, France}
 \affiliation{Dipartimento di Fisica,
Universit\`a di Trento and CNR-INFM BEC Center, I-38050 Povo,
Trento, Italy}

\begin{abstract}
We propose a physical system allowing one to experimentally
observe the distribution of the complex zeros of a random
polynomial. We consider a degenerate, rotating, quasi-ideal atomic
Bose gas prepared in the lowest Landau level. Thermal fluctuations
provide the randomness of the bosonic field and of the locations
of the vortex cores. These vortices can be mapped to zeros of
random polynomials, and observed in the density profile of the
gas.
\end{abstract}


\maketitle

An important field of study in theoretical statistical physics
concerns the properties of the roots of random polynomials
\cite{gen_random_poly,Hannay}. Of particular interest is the
so-called Weyl polynomial for the complex variable $\zeta$:
\begin{equation}
P(\zeta) = \sum_{m=0}^{m_{\rm max}} a_m \frac{\zeta^m}{\sqrt{m!}}
\label{eq:ranpoly}
\end{equation}
where the $a_m$'s are independent random complex numbers with the
same Gaussian probability distribution. The roots of $P$ in the
complex plane can be mapped to a two-dimensional (2D) gas of
particles with repulsive interactions. They are spatially
antibunched, and have a uniform mean density in the large $m_{\rm
max}$ limit \cite{Leboeuf}.

Although the statistical properties of the roots of the Weyl
polynomial have been well studied theoretically, no physical
system has allowed yet to observe them directly. The goal of this
Letter is to show that a 2D rotating ideal Bose gas is a well
suited system for this observation. The positions of the vortices
appearing in the gas can be mapped to the zeroes of the random
polynomial describing the atomic state. More precisely, the gas is
harmonically trapped and it rotates at a frequency close to the
trapping frequency so that it is `frozen' in the lowest Landau
level (LLL) \cite{Rokhsar,Ho}. Its finite temperature $T$ ensures
that several vortices are present in the region where atomic
density is significant, and thermal fluctuations provide the
randomness of the vortex locations in different realizations of
the experiment. Such an experiment is not unrealistic: a 2D atomic
Bose gas in the LLL has recently been produced \cite{expt_lll},
and the use of a Fano-Feshbach scattering resonance allows to
nearly cancel the interactions between ultra-cold atoms
\cite{Feshbach}. This regime of an ideal gas with large thermal
fluctuations dramatically differs from the well studied case of a
rotating and interacting Bose-Einstein condensate in the LLL at
$T=0$, where the vortices are known to form an Abrikosov lattice
\cite{Rokhsar,Ho,expt_lll}.

The ideal gas in our model is confined in a harmonic trap, with
oscillation frequency $\omega$ in the $xy$ plane and $\omega_z$
along $z$. The confinement along $z$ is assumed to be strong,
$k_{\rm B} T \ll \hbar \omega_z$, so that the $z$ degree of
freedom is frozen and the gas is kinematically 2D. We also assume
that some angular momentum has been transferred to the gas by a
stirring procedure \cite{Jean_vortex}, so that thermal equilibrium
is reached in a frame rotating at frequency $\Omega$ around $z$.
$\Omega$ is chosen close to the trapping frequency $\omega$:
 \begin{equation}
\omega-\Omega \ll \omega\ ,
 \end{equation}
which is experimentally realistic since the value $(\omega -
\Omega)/\omega =0.01$ has already been achieved \cite{expt_lll}.
We also assume that the gas is cooled to a low enough temperature,
$k_{\rm B} T \ll 2 \hbar \omega$, so that the relevant single
particle states are linear combinations of the LLL eigenmodes:
\cite{Rokhsar,Ho}
\begin{equation}
\phi_m(x,y) = \frac{\zeta^m}{\sqrt{\pi m!}}\, e^{-\zeta \zeta^*/2}
\ , \qquad \zeta=x+iy\ .
\end{equation}
Here  $a_{\rm ho}=\sqrt{\hbar/M\omega}$ (where $M$ is the atomic
mass) is taken as the unit of length. The mode energy $\epsilon_m
= m\hbar(\omega-\Omega)$ depends on a single quantum number $m\geq
0$ so that thermodynamically the gas is effectively 1D.

The relevant quantity in our study is the density of vortices,
which we will define in relation with the complex classical field
$\psi (\mathbf{r})$ describing the state of the gas. This
classical field represents not only the condensate (atoms in the
vortex-free mode $\phi_0$) but also all superimposed thermal
fluctuations. We obtain $\psi$ using the expression for the
many-body density operator $\hat \sigma$ of the ideal gas in
thermal equilibrium in the grand canonical ensemble \cite{Zoller}:
\begin{equation}
\hat{\sigma} = \int {\cal D}\psi \, P(\{\psi\}) \,
|\mathrm{coh}:\psi\rangle \langle \mathrm{coh}:\psi |\ .
\label{eq:Glauber}
\end{equation}
In this expression $\hat \sigma$ is a statistical mixture of
Glauber coherent states $|\mathrm{coh}:\psi\rangle$, with positive
weights $P(\{\psi\})$ (the so-called Glauber P distribution) given
by the Gaussian functional specified below. A given realization of
the experiment can then be viewed as a random draw of the atomic
field state $\psi$, with the probability law $P(\{\psi\})$.

The stochastic nature of $\psi$ is simple to characterize by
expanding it on the eigenmodes $\phi_m$:
\begin{equation}
\psi(\mathbf{r}) = \sum_{m\geq 0} a_m \phi_m(\mathbf{r}).
\end{equation}
The $a_m$'s are complex, statistically independent random numbers,
with a Gaussian law:
\begin{equation}
P(\{\psi\}) \propto \prod_{m\geq 0} e^{-|a_m|^2/n_m}
\end{equation}
where $n_m=[\exp(\beta(\epsilon_m-\mu)) -1]^{-1}$ is the mean
occupation number of mode $m$. Here $\beta=1/k_{\rm B}T$ and $\mu$
is the chemical potential. This provides a clear link with the
random polynomial of Eq.(\ref{eq:ranpoly}) when several $n_m$ have
similar values:
\begin{equation}
\psi(\mathbf{r}) = f(\zeta) \frac{e^{-\zeta
\zeta^*/2}}{\sqrt{\pi}}\ , \qquad f(\zeta)=\sum_m a_m
\frac{\zeta^m}{\sqrt{m!}}\ .
 \label{eq:psi_f}
\end{equation}
When $f(\zeta)$ is factorized as $f(\zeta)\propto \prod_i
(\zeta-\zeta_i)$, each root $\zeta_i$ corresponds to the location
of a positively charged vortex in the field $\psi$. Since having a
multiple root is a zero measure event, these vortices are of
charge unity. Note that $\psi$ results from the interference of a
large number of macroscopically populated field modes, reminiscent
of the interference of independent condensates
\cite{Ketterle_interf,Dalibard_many_bec}.

The standard case where all the $a_m$'s have the same variance
corresponds in our model to all $\epsilon_m$'s being equal, i.e.
$\Omega=\omega$. The average vortex density is then uniform, $\bar
\rho_v =1/\pi$ and the average pair distribution function $
\rho_2(\mathbf{r}-\mathbf{u}/2,\mathbf{r}+\mathbf{u}/2)$, which
depends only on the relative distance $u$, can be calculated
analytically \cite{Hannay}. However this case can not be achieved
experimentally since for $\Omega=\omega$, the centrifugal force
exactly balances the trapping force and the gas is not confined
anymore. In a realistic model one must address the case
$\Omega<\omega$, for which the trapping force overcomes the
centrifugal one. The statistics of the roots then do not coincide
with the standard results of the literature, and we must perform a
study of this specific model.

\begin{figure}[htb]
\begin{tabular}{ccc}
\includegraphics[height=2.75cm,clip]{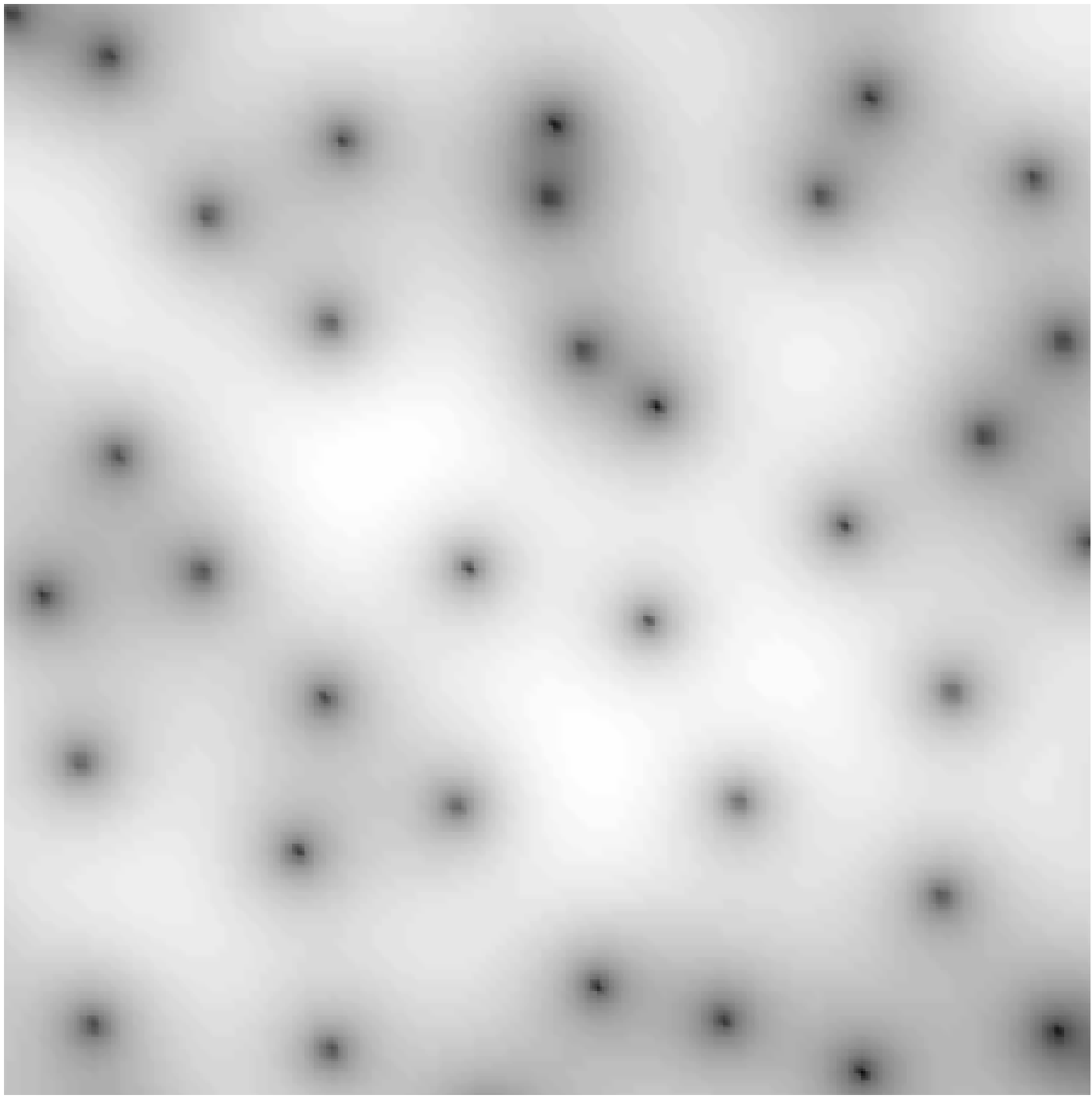} &
\includegraphics[height=2.75cm,clip]{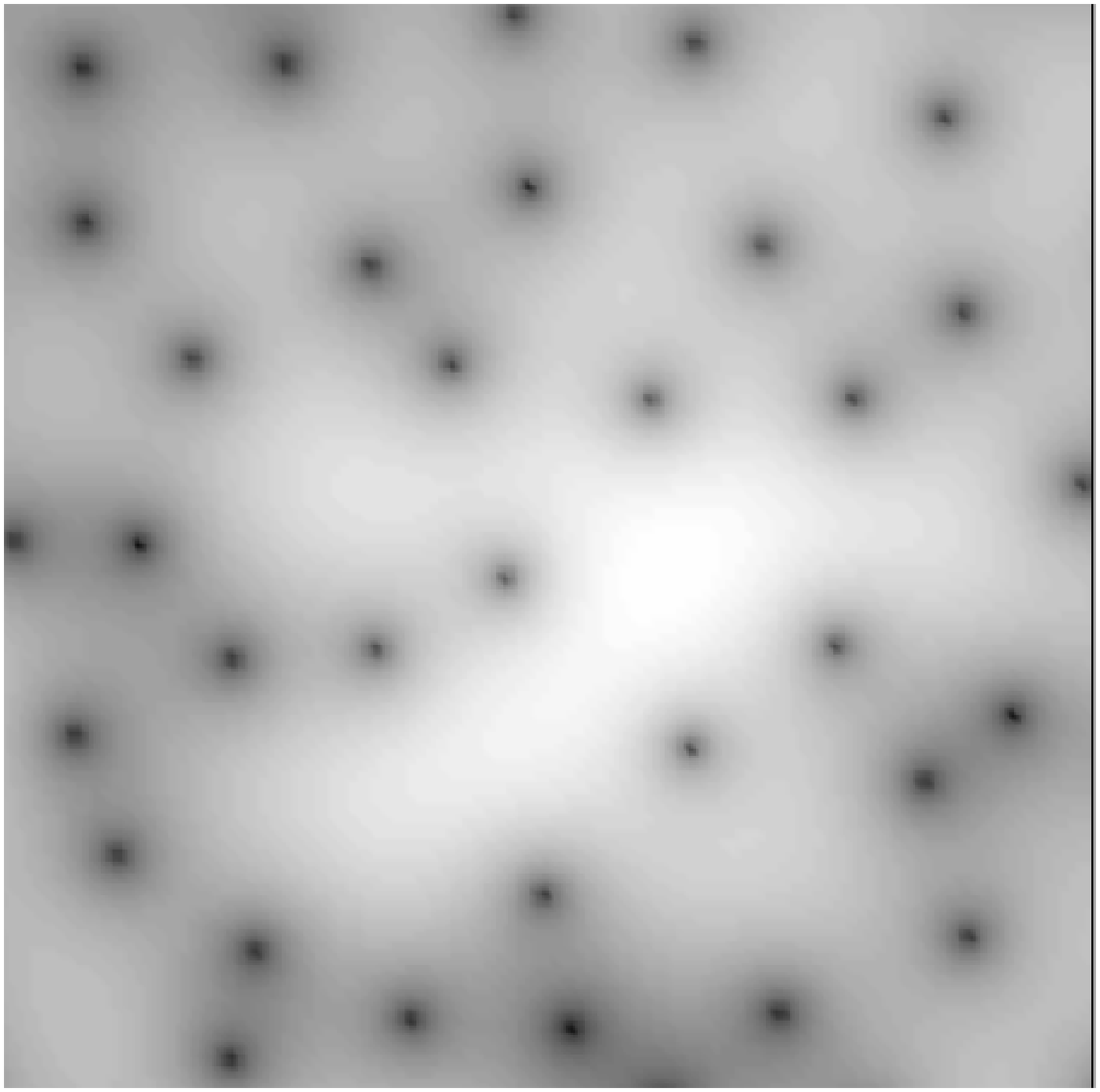} &
\includegraphics[height=2.75cm,clip]{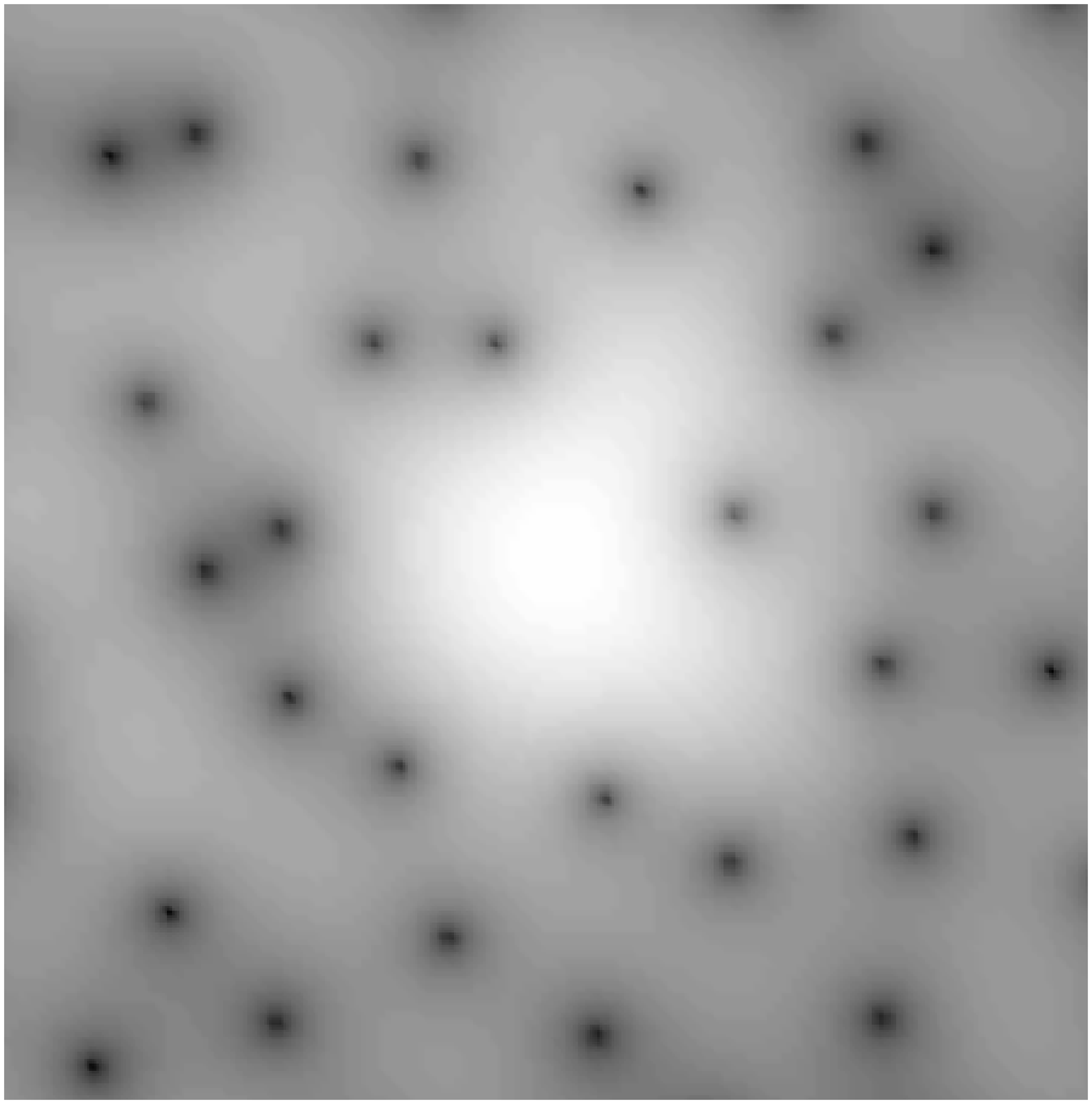} \\
\includegraphics[height=2.75cm,clip]{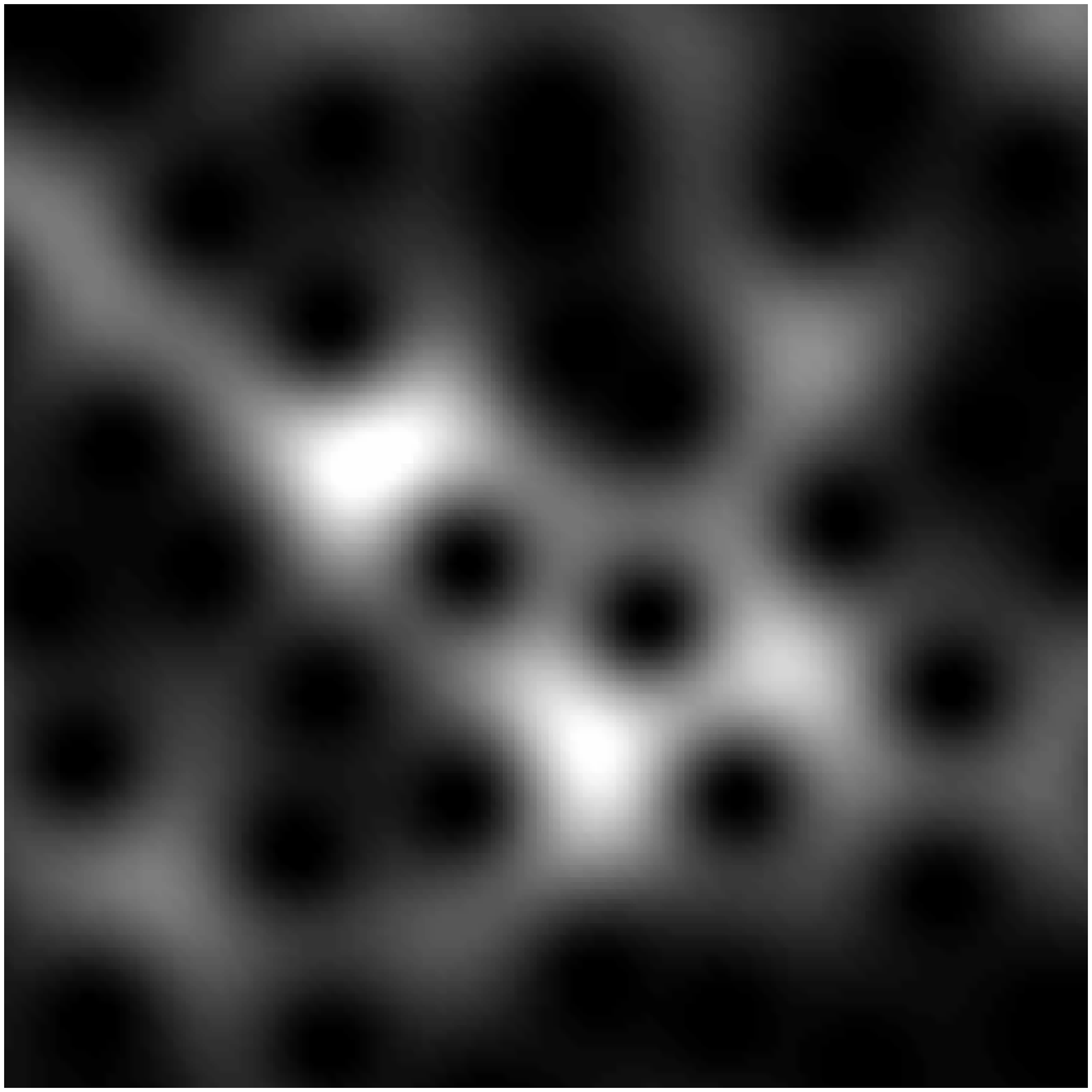} &
\includegraphics[height=2.75cm,clip]{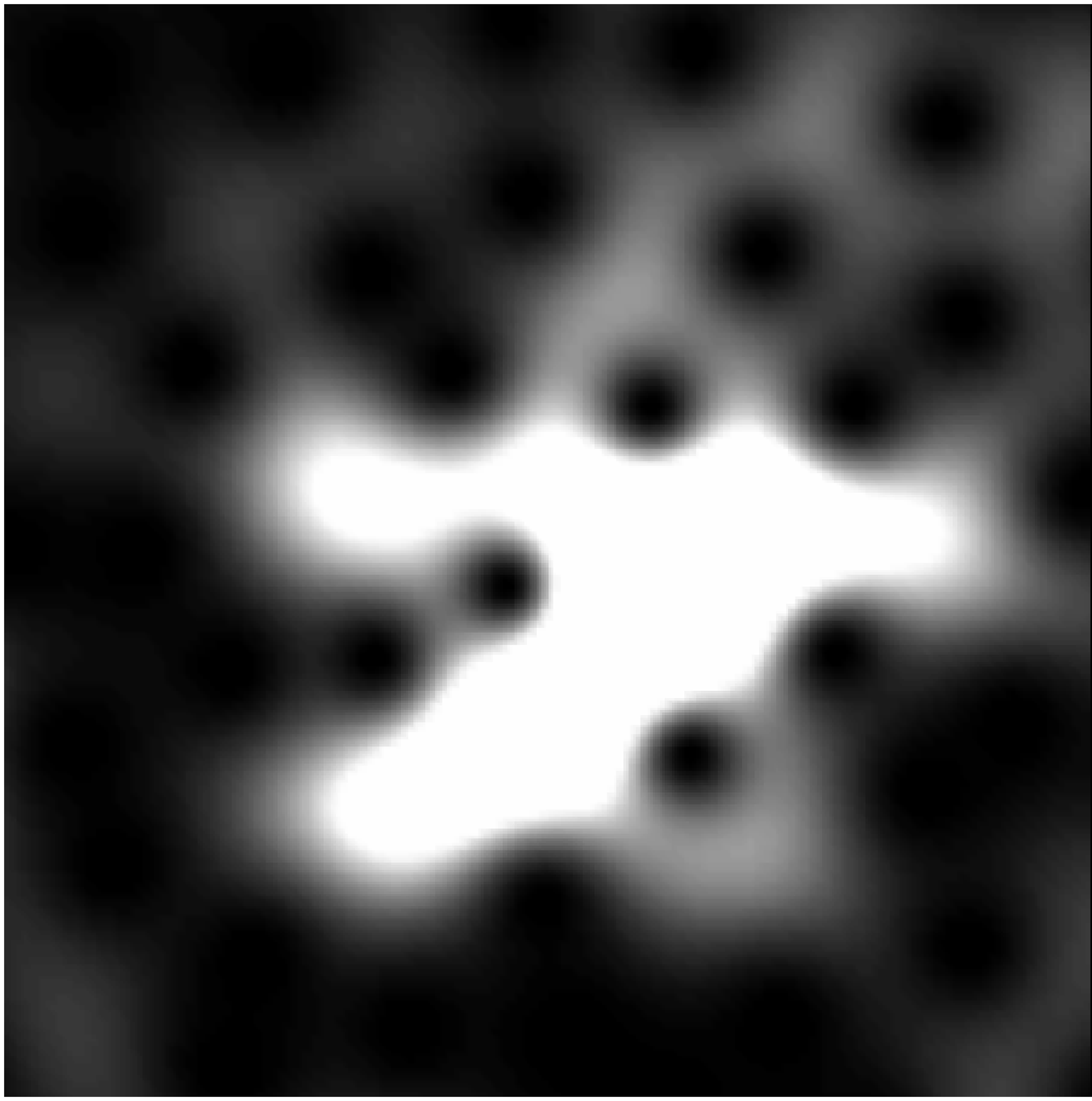} &
\includegraphics[height=2.75cm,clip]{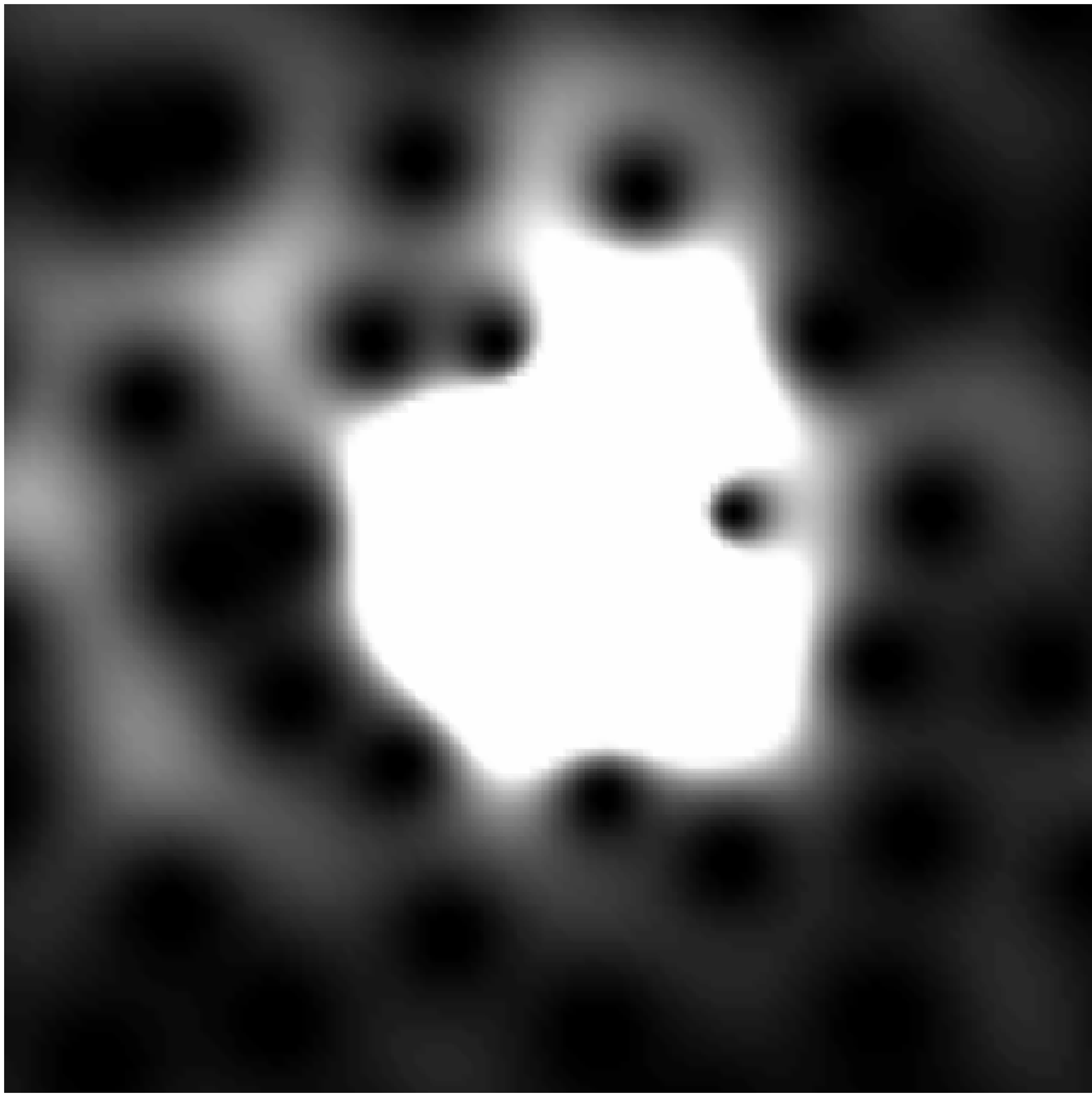} \\
\includegraphics[height=2.75cm,clip]{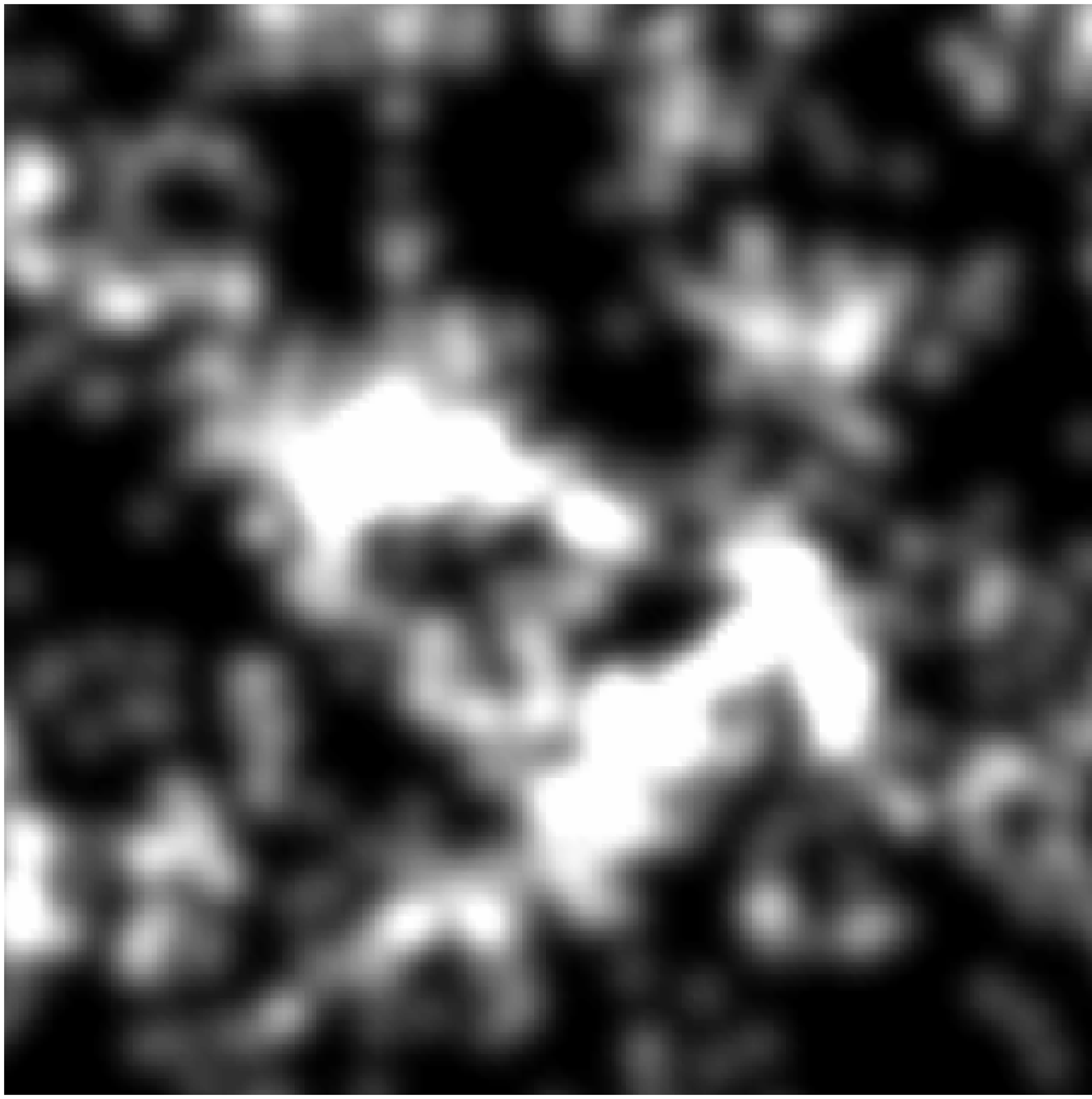} &
\includegraphics[height=2.75cm,clip]{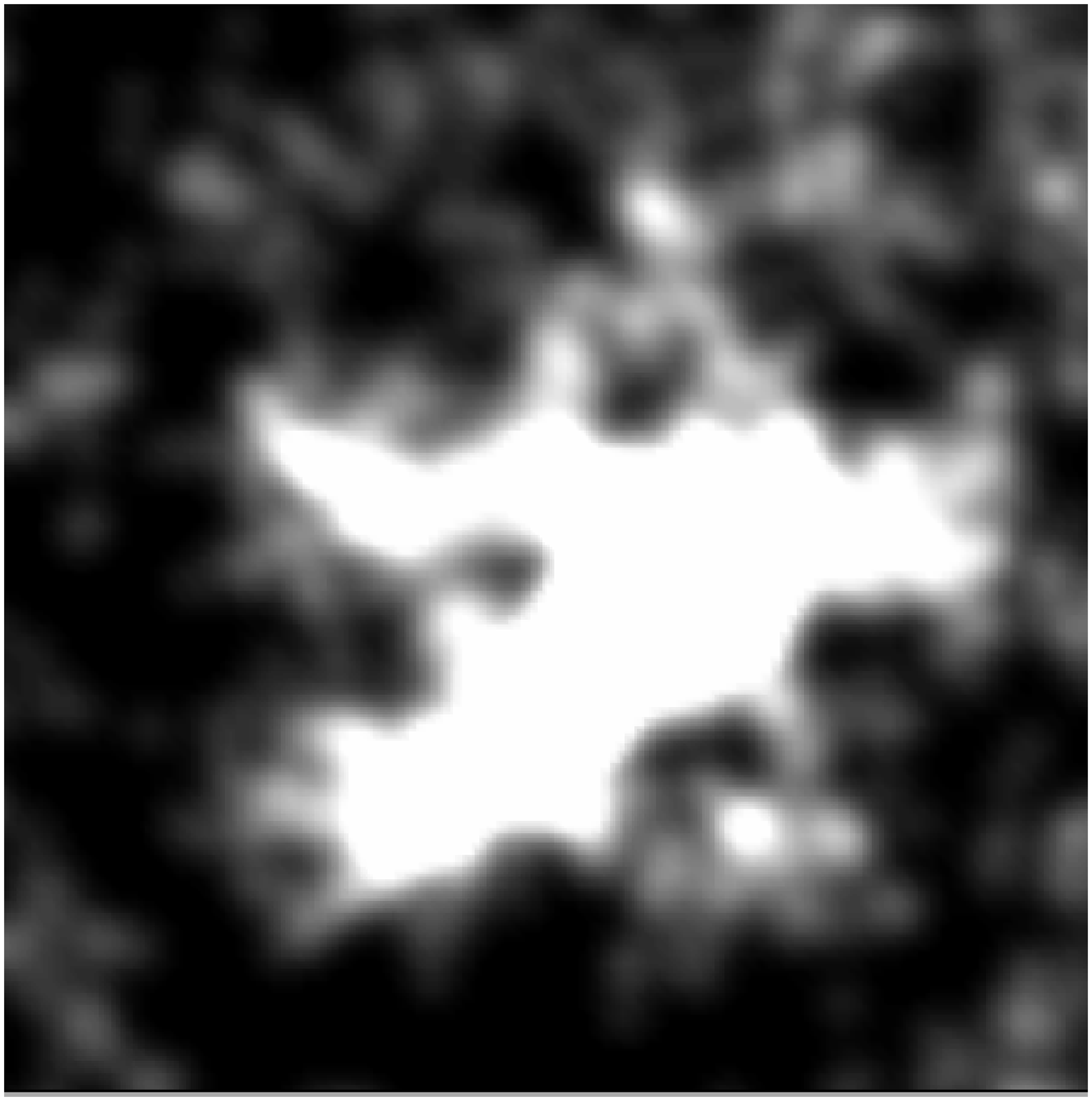} &
\includegraphics[height=2.75cm,clip]{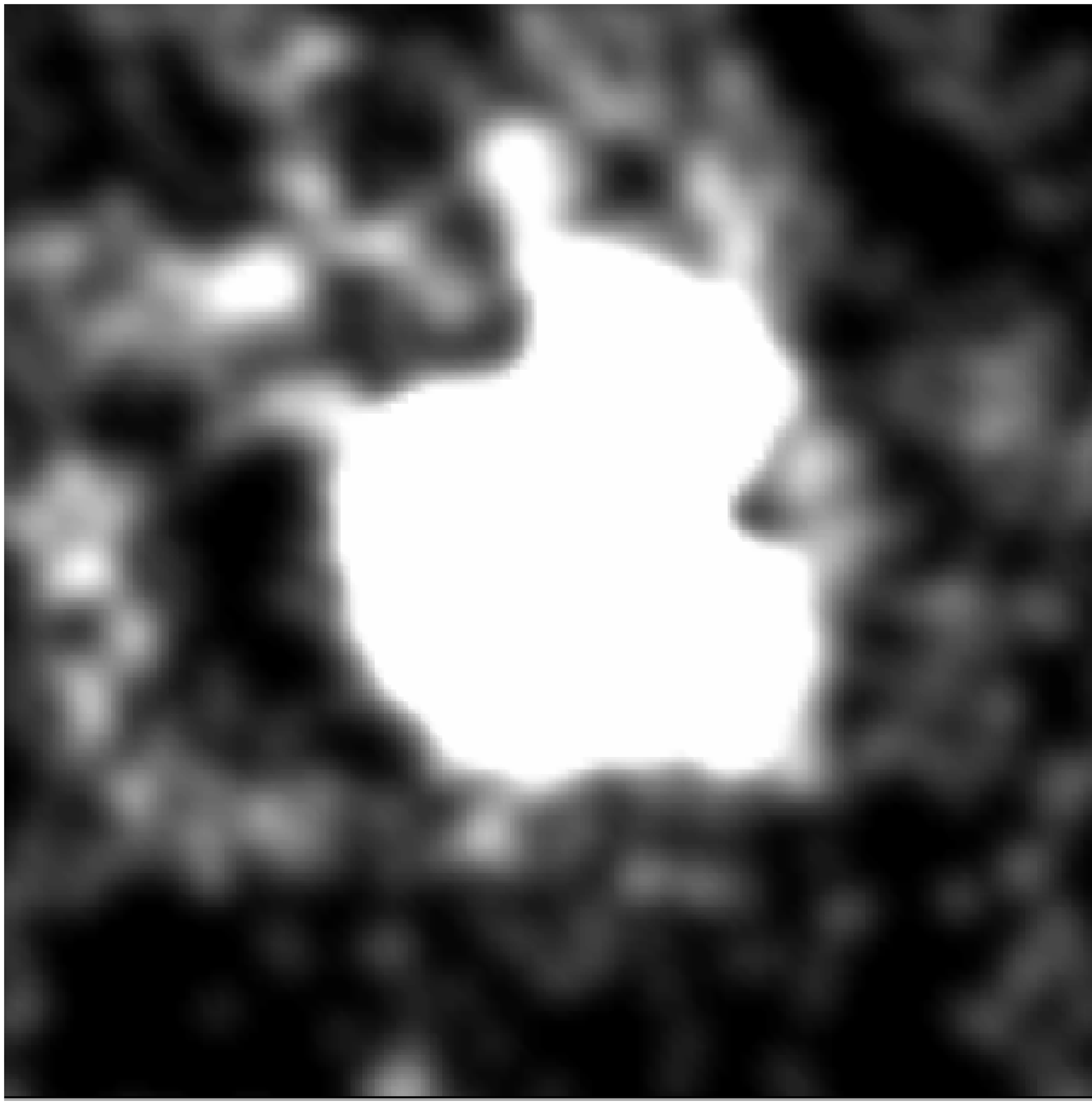} \\
\end{tabular}
\caption{Density plots for three randomly generated fields $\psi$,
using the grand canonical ensemble for $k_B T = 500\, \epsilon_1$
($\epsilon_1=\hbar(\omega-\Omega)$). Left column:
$\mu=-10\,\epsilon_1$, mean number of particles $N=1986$; middle
column: $\mu=-\epsilon_1$, $N=3396$; right column:
$\mu=-0.1\,\epsilon_1$, $N=8320$. First line:
$\ln(|\psi(\mathbf{r})|^2)$. Second line: $|\psi(\mathbf{r})|^2$.
Third line: simulation of a real experiment; $N$ random atomic
positions are generated according to the density
$|\psi(\mathbf{r})|^2$, and a 129$\times$129 pixel `camera' image
is produced assuming that each atom produces a Gaussian spot of
$\sigma=0.2 a_{\rm ho}$. The width of each image is $10 a_{\rm
ho}$. Experimentally, a time of flight technique can be used to
magnify these images, since the LLL modes are self-similar under a
ballistic expansion. } \label{fig:MC}
\end{figure}

To provide an intuitive understanding on how the roots of
$f(\zeta)$ are distributed, we first show in figure \ref{fig:MC}
numerical results for a randomly generated field $\psi$. We take
$k_B T = 500 \,\hbar(\omega-\Omega)$, which is compatible with the
condition $k_{\rm B}T\ll 2\hbar \omega$ for the experimentally
realistic value $\Omega=0.999\,\omega$. On the first and second
lines of figure \ref{fig:MC}, we show plots of $\ln(|\psi|^2)$ and
$|\psi|^2$, respectively, for 3 values of $\mu$. In all cases, the
roots locations are clearly visible.

We now turn to an analytic study of the problem and we express the
algebraic density of vortices $\rho_v(\mathbf{r})$ in terms of
$\psi$ and its derivatives \cite{Halperin}. A helpful
simplification is to eliminate the Gaussian factor in
Eq.(\ref{eq:psi_f}), which does not change the vortex locations,
and use $f(\zeta)$ rather than $\psi(\mathbf{r})$ as a random
field. Using $\partial_y f(\zeta)=i\partial_x f(\zeta)=i
f'(\zeta)$, we obtain:
\begin{equation}
\rho_v(\mathbf{r})= |f'|^2\;\delta^{(2)}[f(\zeta)] \,
\end{equation}
making it obvious that all vortices have a positive charge in the
LLL.

\begin{figure}[tb]
\begin{center}
\includegraphics[height=6.7cm,clip]{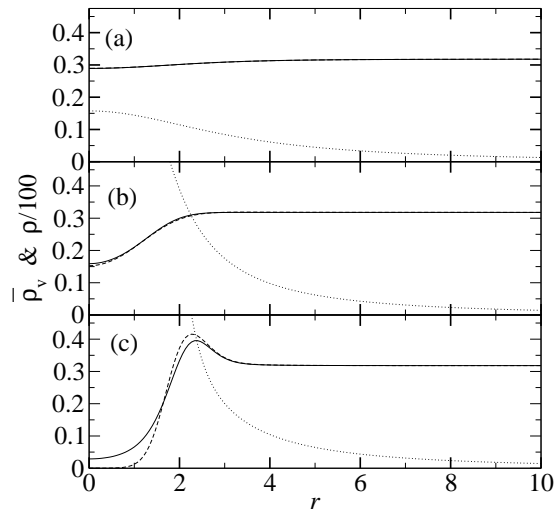}
\end{center}
\vskip -5mm \caption{Vortex ($\bar \rho_v$) and atomic ($\rho$)
densities. Solid line: result for $\bar \rho_v$ in the grand
canonical ensemble. Dashed line: result for $\bar \rho_v$ in the
canonical ensemble with the same mean number of particles $N$.
Dotted line: result for $\rho/100$ in the grand canonical
ensemble. $k_B T = 500 \,\epsilon_1$ with
$\epsilon_1=\hbar(\omega-\Omega)$. (a) $\mu=-10\, \epsilon_1$
($N=1986$). (b) $\mu=-\epsilon_1$ ($N=3396$). (c) $\mu=-0.1\,
\epsilon_1$ ($N=8320$). The unit length is $a_{\rm ho}$.}
\label{fig:rhov}
\end{figure}

The expectation value of $\rho_v$ gives the average vortex density
$\bar \rho_v$, the correlation function $\langle
{\rho}_v(\mathbf{r}-\mathbf{u}/2)
{\rho}_v(\mathbf{r}+\mathbf{u}/2)\rangle$ gives the vortex pair
distribution function $\rho_2$, etc. Using properties of Gaussian
statistics, these quantities can be expressed in terms of
expectation values of products of two fields, such as $f$, $f^*$,
or their derivatives with respect to $\zeta$ \cite{Hannay}. The
explicit result for $\bar \rho_v$ is:
\begin{equation}
{\bar\rho}_v(\mathbf{r})=\frac{\langle f'^{*} f'\rangle} {\pi
\langle f^* f\rangle} -\frac{\langle f^* f'\rangle \, \langle f
f'^{*} \rangle } {\pi \langle f^* f\rangle^2}.
\label{eq:meandensv}
\end{equation}
The expectation values $\langle \ldots\rangle$ are readily
calculated using $\langle a_m^* a_{m'}\rangle = n_m
\delta_{m,m'}$. Figure~\ref{fig:rhov} gives the variation of $\bar
\rho_v$ with $r$, for the temperature and the three values of the
chemical potential used in figure~\ref{fig:MC}.

The general expression for the pair distribution function $\rho_2$
is rather involved, and we give it only for small relative
distances $u$. It vanishes quadratically with $u$, showing the
effective repulsion between the vortices:
\begin{equation}
\rho_2(\mathbf{r}-\frac{\mathbf{u}}{2},\mathbf{r}+\frac{\mathbf{u}}{2})
= C(\mathbf{r}) \frac{u^2}{8\pi^2} + O(u^3)\ , \label{eq:quad}
\end{equation}
where $C$ is a function of  $f$, $f^*$, and their first and second
derivatives \cite{valC}.

We find that the experimentally relevant situation $\Omega<\omega$
leads to results similar to those of the standard random
polynomial theory (formally corresponding to $\Omega=\omega$) in
the vicinity of the trap center when the two following conditions
are fulfilled. Firstly, many eigenmodes have to be thermally
populated:
\begin{equation}
k_B T \gg \hbar (\omega-\Omega)\ . \label{eq:high_temp}
\end{equation}
Secondly, the low energy modes must have comparable populations,
imposing a negative chemical potential $|\mu|\gg \hbar
(\omega-\Omega)$. This is the situation depicted in the left
column of figure~\ref{fig:MC} and in figure \ref{fig:rhov}a.

If $|\mu|$ is reduced for a fixed $T$, the mean number of atoms
increases and an imbalance appears among the populations of the
low energy modes. The vortex density is then depleted near the
trap center, as it can be seen in the middle and right columns of
figure~\ref{fig:MC}, and in figure \ref{fig:rhov}b,c. Eventually,
for $|\mu|\ll \hbar(\omega-\Omega)$, the total population of the
modes $m\neq 0$ saturates and a condensate forms in the mode
$m=0$. This condensate has no vortex and it expels the thermal
vortices, which accumulate in a corona with a density locally
exceeding $1/\pi$ (figure~\ref{fig:rhov}c).

The situation at the onset of condensation is relevant
experimentally since it is tempting to increase the number of
atoms in order to improve the experimental signal. To extend
quantitatively our analysis to this regime, we must replace the
grand canonical ensemble, which presents non physical, large
fluctuations of the total particle number when a condensate is
present, by the canonical ensemble. We then adapt our definition
of the vortex distribution by projecting Eq.(\ref{eq:Glauber}) on
a subspace with a fixed total number of particles $N$, resulting
in the following probability distribution of the field $\psi$:
\begin{equation}
P_c(\{\psi\}) \propto P(\{\psi\}) \frac{||\psi||^{2N}}{N!}
e^{-||\psi||^2} \label{eq:rho_can}
\end{equation}
where $||\psi||^2=\int |\psi|^2$. The key change with respect to
the grand canonical ensemble is that the probability distribution
of the field is no longer Gaussian, showing that the proposed
physical system opens a new class of problems. We evaluate
numerically the vortex density using a generating function
technique \cite{technical}. In the trap center, one can also
construct an exact mapping to a solvable problem, the calculation
of the canonical partition function and occupation numbers for a
1D ideal Bose gas in a harmonic trap \cite{Olshanii_Herzog}. This
leads to the exact expression:
\begin{equation}
\bar \rho_v^c(\mathbf{r}=\mathbf{0})= \frac{1-
e^{-\beta\epsilon_1}}{\pi}\left[\frac{1}{e^{\beta\epsilon_1}-1}-n_{1}^{c}
\right]
\end{equation}
where $n_1^{c}$ is the mean number of particles in the first
excited mode $m=1$ (with $\epsilon_1=\hbar(\omega-\Omega)$) in the
canonical ensemble \cite{n1c}. As shown in figure \ref{fig:rhov}c,
the correct canonical result drops to a much smaller value than
the incorrect grand canonical result in the trap center. In the
large $N$ limit, one can show that $\bar{\rho}_v^c(\mathbf{0})\sim
N \exp[-(N+1)\beta\epsilon_1]/\pi$ whereas the grand canonical
prediction tends to zero only as $1/N$. Note that
$\bar\rho_v^c(\mathbf{0})$ drops with $N$ similarly to the
probability $\exp(-N\beta\epsilon_1)$ of having an empty
condensate mode, which is a natural condition to have a vortex in
$\mathbf{r}=\mathbf{0}$.

How to observe the vortex density in practice? In current
experiments with rotating interacting condensates, one measures
the positions of the particles and the vortices appear as holes in
the density profile \cite{Jean_vortex}. This strategy can be used
in regions of space where the mean density of particles $\rho$
greatly exceeds the density of vortices $\bar \rho_v$, here $ \sim
1/\pi$. To each vortex embedded in such a high density region will
correspond a clearly identifiable hole in the particle
distribution. The maximal density of the non-condensed fraction of
the gas is $\sim k_B T/[\pi\,\hbar(\omega-\Omega)]$ close to the
trap center, obtained for $|\mu|\ll \hbar(\omega-\Omega)$. Then
$\rho\gg \bar \rho_v$ when Eq.(\ref{eq:high_temp}) is satisfied.
To indicate up to what distance from the trap center this
condition holds, we have also plotted $\rho(r)/100$ in figure
\ref{fig:rhov}.

Conversely, do all the holes embedded in high density regions
correspond to vortices? They may in principle correspond to a
local minimum of $|\psi|^2$, where the field assumes a small but
non zero value. Such `spurious' local minima can in general form
in a non-condensed ideal Bose gas, which is subject to large
density fluctuations due to the thermal bunching effect of the
bosons. However, we can show that this phenomenon is absent in the
LLL. We Taylor-expand the field $\psi$ to second order in
$\mathbf{u}\equiv \mathbf{r}-\mathbf{r_0}$ around the location
$\mathbf{r_0}$ of a stationary point of $|\psi|^2$, assuming that
$\psi(\mathbf{r_0})\neq 0$:
\begin{equation}
|\psi(\mathbf{r})|^2=|\psi(\mathbf{r_0})|^2\,
\left[1+ \mathbf{u}\cdot M \mathbf{u} + O(u^3)\right]
\end{equation}
where the 2$\times$2 matrix $M$ is real symmetric.
One then finds that the trace of $M$ is
\begin{equation}
\mathrm{Tr}\, M = \mathrm{Re}\, \frac{\Delta\psi}{\psi}
+\left|\frac{\mathbf{grad}\,\psi}{\psi}\right|^2 = -2
\end{equation}
where we used Eq.(\ref{eq:psi_f}) and
$\mathbf{grad}(|\psi|^2)=\mathbf{0}$ to get the last identity.
Since its trace is $<0$, $M$ cannot be positive and $|\psi|^2$
cannot have a local minimum at $\mathbf{r}_0$.

We exemplify the above discussion by a Monte Carlo simulation of a
real experiment for the parameters of figure \ref{fig:MC}. The
results are shown on the last line of figure~\ref{fig:MC}.
Starting from the random fields $\psi$ previously generated, we
have produced these images by generating random positions of
particles according to the distribution $|\psi|^2$, and by
mimicking the finite imaging resolution of a real experiment.
Local minima of the density are visible, and can be checked on the
images of the first and second lines to correspond to vortices.
The visibility of the vortex pattern could be improved by
increasing further the atom density. For a given ratio $k_{\rm B}
T/\hbar \omega$, this can be achieved by rotating the gas even
faster, so that $(\omega-\Omega)/\omega$ decreases.

In real life, there are interactions between the particles,
characterized by the 3D $s$-wave scattering length $a$. The
interaction potential in the LLL is then modeled by the
pseudo-potential $g\,\delta^{(2)}(\mathbf{r}_1-\mathbf{r}_2)$,
with $g=2\sqrt{2\pi}\hbar\omega a/a_z$, where
$a_z=\sqrt{\hbar/M\omega_z}$. We now derive a condition on $a$ for
the interactions to be negligible, in calculating to first order
in $g$ the effect of the interactions on the probability
distribution of the vortices. Focusing on the quasi-homogeneous
regime $|\mu|\gg \hbar(\omega -\Omega)$, we set $\Omega=\omega$.
The unperturbed occupation numbers then all have the same value
$n_0$. The vortex density being uniform, even in presence of
interactions, we consider the first order correction in $g$ to the
vortex pair distribution function $\rho_2(\mathbf{u})$. Its
expression can be simplified in the degenerate regime $n_0\gg 1$:
\begin{equation}
\frac{\delta \rho_2 (\mathbf{u})}{\rho_2^{(0)}(\mathbf{u})} \simeq
-\frac{\beta g}{2}\int d^2\mathbf{r}\, \left[\frac{\langle
\rho_v(\mathbf{0})\rho_v(\mathbf{u})\,|\psi(\mathbf{r})|^4
\rangle}{\rho_2^{(0)}(\mathbf{u})} -\langle
|\psi(\mathbf{r})|^4\rangle\right], \label{eq:delta_rho2}
\end{equation}
where the average is taken over the unperturbed distribution. The
right hand side of Eq.(\ref{eq:delta_rho2}) can be expressed
analytically. It is maximal (in absolute value) in $u=0$:
\begin{equation}
\lim_{u\rightarrow 0} \frac{\delta\rho_2(\mathbf{u})}{\rho_2^{(0)
}(\mathbf{u})}= -\frac{3\beta g n_0^2}{16\pi} \propto \frac{g
\rho}{|\mu|},
\end{equation}
whereas it  drops as $u^4 e^{-u^2/2}$ at large distances.
Interactions will then play a negligible role in the vortex
distribution if the mean field energy $g\rho$ is much smaller than
$|\mu|$. For the realistic numbers $k_B T=\hbar\omega/2$,
$n_0=100$ and $a_z=1\,\mu\mathrm{m}$, the condition for negligible
interactions is $|a| < 0.2\, \mathrm{nm}$. The scattering length
can indeed be tuned in practice to such a low value using a
Fano-Feshbach resonance. This tuning should be done only during
the last fraction of the experimental sequence, after having taken
advantage of a larger value of $a$ to reach thermal equilibrium in
a reasonable time.

To summarize, the rotating quasi-ideal Bose gas is a promising
system to implement in practice the concepts developed in the
theory of random polynomials. It also raises novel questions such
as the influence of the non-Gaussian statistics for the polynomial
coefficients when a condensate is present. Finally we emphasize
that our disordered vortex pattern appears in a non-superfluid
system. An interesting future line of research is the transition
to an ordered Abrikosov lattice when interactions are increased
and the system becomes superfluid.

We thank Baptiste Battelier and Iacopo Carusotto for useful
discussions. S. Stock acknowledges support from the
Studienstiftung des deutschen Volkes and the DAAD, and Z.
Hadzibabic from the EU (Contract No. MIF1-CT-2005-00793). This
work is partially supported by R\'egion Ile de France and ACI
Nanoscience. Laboratoire Kastler Brossel is a research unit of
Ecole normale sup\'erieure and Universit\'e Paris 6, associated to
CNRS.


\begin{thebibliography}{99}
\bibitem{valC}
We found $C=B^2/A^3$, with $A=\langle 0,0\rangle \langle 1,1\rangle - |\langle 1,0\rangle|^2$
and $B=-\langle 2,2\rangle A- [\langle 2,1\rangle \langle 1,0\rangle \langle 0,2\rangle+\mbox{c.c.}]
+|\langle 2,1\rangle|^2 \langle 0,0\rangle + |\langle 2,0\rangle|^2 \langle 1,1\rangle$, using
the notation $\langle p,q\rangle= \langle f^{(p)*}f^{(q)}\rangle$.
\bibitem{technical}
One defines the generating function
$B(\theta) = \langle \exp[-||\psi||^2 (1+e^{i\theta})]\rangle_P$.
Then one computes the coefficient of $B$
on the Fourier component $e^{iN\theta}$, which gives the normalization factor
of $P_c(\{\psi\})$.  A similar trick applies
for the vortex density, so that
$\bar{\rho}_{v,c} =
\int_0^{2\pi} d\theta\, \bar{\rho}_{v,\theta} B(\theta)
e^{-iN\theta}/
\int_0^{2\pi} d\theta\, B(\theta) e^{-iN\theta}$
where $\bar{\rho}_{v,\theta}$ is taken from Eq.(\ref{eq:meandensv})
replacing the mean occupation numbers $n_m$
by $[\exp(\beta(\epsilon_m-\mu))+\exp i\theta]^{-1}$.
\bibitem{n1c}
The value $n_{1}^{c}(N)$ for $N$ particles is obtained from
the recursion $n_{1}^{c}(N)=e^{-\beta\epsilon_1} (1-e^{-\beta N \epsilon_1})
[1+n_{1}^{c}(N-1)]$ \cite{Olshanii_Herzog}.
\end{thebibliography}
\end{document}